\documentclass[journal]{IEEEtran}
\ifCLASSINFOpdf
\else
\fi
\usepackage{graphicx}
\usepackage{float}
\usepackage{amssymb}
\usepackage{amsmath}
\usepackage{cite}
\usepackage{multirow}
\usepackage{stfloats}

\begin{document}
\title{Self-supervised Remote Sensing Images Change Detection at Pixel-level}
%
\author{Yuxing~Chen,~\IEEEmembership{~}
	    Lorenzo~Bruzzone,~\IEEEmembership{Fellow,~IEEE}
        
\thanks{Y. Chen and L. Bruzzone are with the Department of Information Engineering and Computer Science, University of Trento, 38122 Trento, Italy (e-mail:chenyuxing16@mails.ucas.ac.cn;lorenzo.bruzzone@unitn.it).}
\thanks{Corresponding author: L. Bruzzone}}

\maketitle

\begin{abstract}
Deep learning techniques have achieved great success in remote sensing image change detection. Most of them are supervised techniques, which usually require large amounts of training data and are limited to a particular application. Self-supervised methods as an unsupervised approach are popularly used to solve this problem and are widely used in unsupervised binary change detection tasks. However, the existing self-supervised methods in change detection are based on pre-tasks or at patch-level, which may be sub-optimal for pixel-wise change detection tasks. Therefore, in this work, a pixel-wise contrastive approach is proposed to overcome this limitation. This is achieved by using contrastive loss in pixel-level features on an unlabeled multi-view setting. In this approach, a Siamese ResUnet is trained to obtain pixel-wise representations and to align features from shifted positive pairs. Meanwhile, vector quantization is used to augment the learned features in two branches. The final binary change map is obtained by subtracting features of one branch from features of the other branch and using the Rosin thresholding method. To overcome the effects of regular seasonal changes in binary change maps, we also used an uncertainty method to enhance the temporal robustness of the proposed approach. Two homogeneous (OSCD and MUDS) datasets and one heterogeneous (California Flood) dataset are used to evaluate the performance of the proposed approach. Results demonstrate improvements in both efficiency and accuracy over the patch-wise multi-view contrastive method.
\end{abstract}


\IEEEpeerreviewmaketitle

\section{Introduction}
\IEEEPARstart{T}{he} basic idea of self-supervised change detection in remote sensing is to align the shared information between multi-view images but reduce the impact of the sensor- and time-related noise.
Multi-temporal remote sensing images can provide time-related information, such as the land-cover changes, but they also contain seasonal noises.
Except for the seasonal noise, there are also sensor-related noises within cross-sensor image pairs.
In this context, there are many benefits for self-supervised learning in multi-view remote sensing image change detection.
On the one hand, self-supervised learning methods can obviate the need for labels in change detection tasks and the learned features can be used in downstream tasks.
On the other hand, self-supervised methods, especially contrastive learning methods, can be used in multi-view scenarios including multi-sensor and multi-resolution image pairs.

Recently, self-supervised learning \cite{goodfellow2014generative,masci2011stacked,tian2019contrastive,geng2017classification,niu2018conditional} is treated as a promising candidate for obtaining meaningful representations and overcoming both types of noises.
The intuitive idea is to reconstruct the counterpart of multi-views using generative models.
Nevertheless, some studies have shown that such generative models overly focus on pixels rather than on abstract feature representations \cite{liu2020self}.
Recent researches in contrastive learning \cite{tian2019contrastive,he2020momentum,chen2020simple} encourage the network to learn more interpretable and meaningful feature representations from multi-view images, where they outperformed the generative counterparts.
Application of such algorithm, in working with multi-view remote sensing image change detection, have been studied in several works \cite{chen2021self,leenstra2021self} and they show very promising results.
However, most of these works focus on an image-level classification or pre-defined tasks.
And, the patch-based algorithm makes it very computationally expensive.
How to perform contrastive learning at the pixel-wise change detection is a problem that until now has been relatively unexplored.
Follow the smart architecture design of the non-contrastive self-supervised method (such as BYOL\cite{grill2020bootstrap} and SimSiam \cite{chen2020exploring}), some works \cite{chen2021shift, xie2020propagate} implemented the pixel-wise feature representation learning by introducing a pixel-level shift equivariance task.
Although those methods obtain competitive performance in the homogeneous image representation learning, they cannot be applied to heterogeneous images directly as the limitation of Siamese network architecture.

In addition, few self-supervised change detection approaches have considered the aleatoric uncertainty of regular seasonal changes for binary change detection tasks.
For example, the feature of cropland with a high uncertainty while the forest is more stable.
This is because the cropland has a significant change along with seasonal changes.
The uncertainty information can help to reduce the impact of such regular changes.
In the CV community, some novel approaches \cite{mi2019training,kendall2017uncertainties} have been proposed to estimate aleatoric uncertainty during the training and inference of models.
For the regression task, models can predict the uncertainty in one forward. 
Most of them require training labels to perform uncertainty estimation during training.
Unfortunately, there are no labels that can be used in such an unsupervised change detection task.

For these reasons, in this work, we propose a pixel-wise self-supervised change detection approach based on contrastive learning.
More precisely, the proposed approach is trained on a pre-task of shift equivariance and each pixel is treated as a single instance in the contrastive loss.
The input images of two branches have a relative shift, and then a reverse operation was performed on the output features for aligning two branches.
Pixels obtained from the same location of multi-view image pairs are called positive pairs.
Negative pairs are obtained from different locations or different batches of multi-view image pairs.
Besides, the vector quantizer is used to augment learned features in two branches.
In the training process, the contrastive loss was adopted to make the features close for positive pairs while making the negative pairs apart.

In addition to the contrastive approach, we also introduce an uncertainty approach to reduce the impact of regular seasonal changes at the second step of the network training.
We have further shown that the uncertainty approach can improve the performance of the binary change maps generated by contrastive learning.
Finally, we demonstrate the effectiveness of our approach on two homogeneous (OSCD and MUDS) and one heterogeneous (California flood) data sets.

In summary, our contributions are the following:
\begin{itemize}
	\item As far as we know, we are the first to apply the pixel-wise contrastive method in remote sensing change detection tasks, and verify it on multi-temporal and multi-sensor datasets.
	\item We propose a new self-supervised change detection approach at pixel level and introduce a simple but useful uncertainty approach in the change detection task to reduce the impact of seasonal changes.
	\item We provide a comparison with the state-of-the-art methods on two types of datasets. Experiment results show that our method obtained comparative results on the OSCD and Flood dataset and the improvement of kappa by 10\% on the MUDS compared to the patch-based contrastive method. Moreover, the proposed approach outperforms the patch-based contrastive method in the efficiency and robustness of water areas. We have further shown the improvements of the uncertainty approach on OSCD and MUDS datasets.
\end{itemize}

The rest of this paper is organized as follows.  
Section II presents the related works of self-supervised change detection in multi-view images and uncertainty estimation.  
Section III introduces the proposed approach by describing the architecture of the network, the considered contrastive loss and the uncertainty approach based on the teacher-student paradigm.  
The experimental results obtained on two types of datasets and the related comparison with the state-of-the-art methods as well as the discussion of results are illustrated in Section IV. 
Finally, Section V draws the conclusions of the paper.

\section{Related Works}
\subsection{Self-supervised Change detection}
In the literature, self-supervised change detection techniques in multi-view remote sensing images can generally be grouped into two categories: generative- and discriminative-based methods.
Generative models \cite{lv2018deep,bergamasco2019unsupervised,kalinicheva2019change,ren2020unsupervised,dong2020self} have been adopted to generate features of multi-temporal or multi-sensor image pairs and detect the changes by an explicit comparison of the generated features.
Liu \textit{et al.} propose a stacked autoencoder to extract the temporal change features of multi-temporal SAR images based on superpixels.
Bergamasco \textit{et al.} \cite{bergamasco2019unsupervised} futher propose to use a multilayer convolutional autoencoder (CAE) for multi-temporal Sentinel-1 images change detetion in a unsupervised fashion.
Besides the autoencoder, generative adversarial networks are also used for change detection tasks.
Gong \textit{et al.} \cite{gong2017generative} treat change detection as a generative learning procedure that connected bi-temporal images and generated the desired change map.
Due to the mis-coregistration between multi-temporal very high resolution (VHR) images, Ren \textit{et al.} \cite{ren2020unsupervised} use the generative adversarial network (GAN) to generate better-coregistered images, and then generate binary change maps by comparing these generated images explicitly.
Dong \textit{et al.} \cite{dong2020self} utilize the GAN's discriminator to differentiate samples from bi-temporal images and transformed bi-temporal images into more consistent feature representations for direct comparison.
Not only for the homogeneous image pairs but generative models are also used for heterogeneous image pairs change detection.
In \cite{luppino2020code}, Luppino \textit{et al.} jointly use domain-specific affinity matrices and autoencoders to align the related pixels from multimodal images.
Niu \textit{et al.} \cite{niu2018conditional} propose the conditional generative adversarial network (cGAN) to translate two heterogeneous images into a single domain for comparison directly.
Liu \textit{et al.} \cite{liu2021unsupervised} futher use the cycle-consistent adversarial networks (CycleGANs) to learn the mapping relation between heterogeneous image pairs.

Similarly, discriminative models, including pre-defined task and contrastive methods, are also used in this domain.
In \cite{leenstra2021self}, Leenstra \textit{et al.} defined two pretext tasks, discriminating overlapping or non-overlapping patches and minimizing the distance between overlapping patches, for feature representation learning.
They further pre-train a discriminative model to extract features from bi-temporal images on these pretext tasks and finetune the trained network to detect changes.
Although pretext tasks were widely used in self-supervised learning, it is not direct for the change detection task.
Contrastive learning is a discriminative self-supervised approach, which compares instance features directly and aims to make similar samples close while making different samples apart.
To achieve this, the InfoNCE \cite{oord2018representation} was widely used in the earlier works \cite{oord2018representation,tian2019contrastive} for the instance-level classification task.
Combined with Siamese networks and momentum encoders, recent methods such as SwAV \cite{caron2020unsupervised}, MoCo\cite{he2020momentum}, and SimCLR \cite{chen2020simple} show comparable results to the state-of-the-art supervised method on several downstream tasks.
An early attempt to use contrastive methods in change detection is \cite{chen2021self}, authors proposed to use contrastive learning in multi-view remote sensing images (including multi-temporal and multi-sensor) change detection directly.
Instead of the pixel-based approach, authors process change detection based on patches.
In \cite{chen2021self}, they also tried to use BYOL for change detection in SAR-optical fusion scenario.
Besides these two methods, Saha \textit{et al.} \cite{saha2021self} integrate contrastive learning, deep clustering, and Siamese network for unsupervised multi-sensor change detection tasks.

\subsection{Uncertainty Estimation}
The uncertainty of results obtained from deep learning networks resulting from two components: aleatoric and epistemic uncertainty \cite{kendall2017uncertainties}.
In more detail, the aleatoric uncertainty is related to the noise in observations, whereas the epistemic uncertainty origin from model parameters.
In \cite{kendall2017uncertainties}, Kendall and Gal introduced an efficient approach to estimate data-dependent aleatoric uncertainty in a predictive manner, which can be learned with a KL loss function.  
In practice, they trained a deep neural network with the output of mean and variance values.
Then the variance value is used as the weight of the distance between mean values and ground truth, and that is used as a regularization term to discourages infinite predictions.
Compared with bayesian methods, this approach is effective and direct and can be completed in just one forward pass.
Therefore, it has been widely used in regression tasks \cite{poggi2020uncertainty,shen2021real,ke2020deep}.
In \cite{ke2020deep}, authors employed the KL loss on the estimated depths and ground truth depth for predicting the depth's uncertainty in a 3D scene reconstruction task, thus improving reconstruction accuracy. 
Besides the supervised fashion, Poggi \textit{et al.} \cite{poggi2020uncertainty} proposed a teacher-student framework to estimate depth's uncertainty in self-supervised monocular depth estimation.
Specifically, they first trained a teacher model in a self-supervised manner and then trained a student model to mimic the distribution sourced from the teacher model.
Finally, they minimized the KL loss function on the estimated depths from the teacher and student models.
Similarly, Shen \textit{et al.} \cite{shen2021real} also straight-forward used a teacher-student paradigm for depth's uncertainty estimation, while the predictive samples are generated from the bayesian teacher using MC-dropout \cite{gal2016dropout}.

\begin{figure*}[pt]
	\centering
	\includegraphics[width=7.0in]{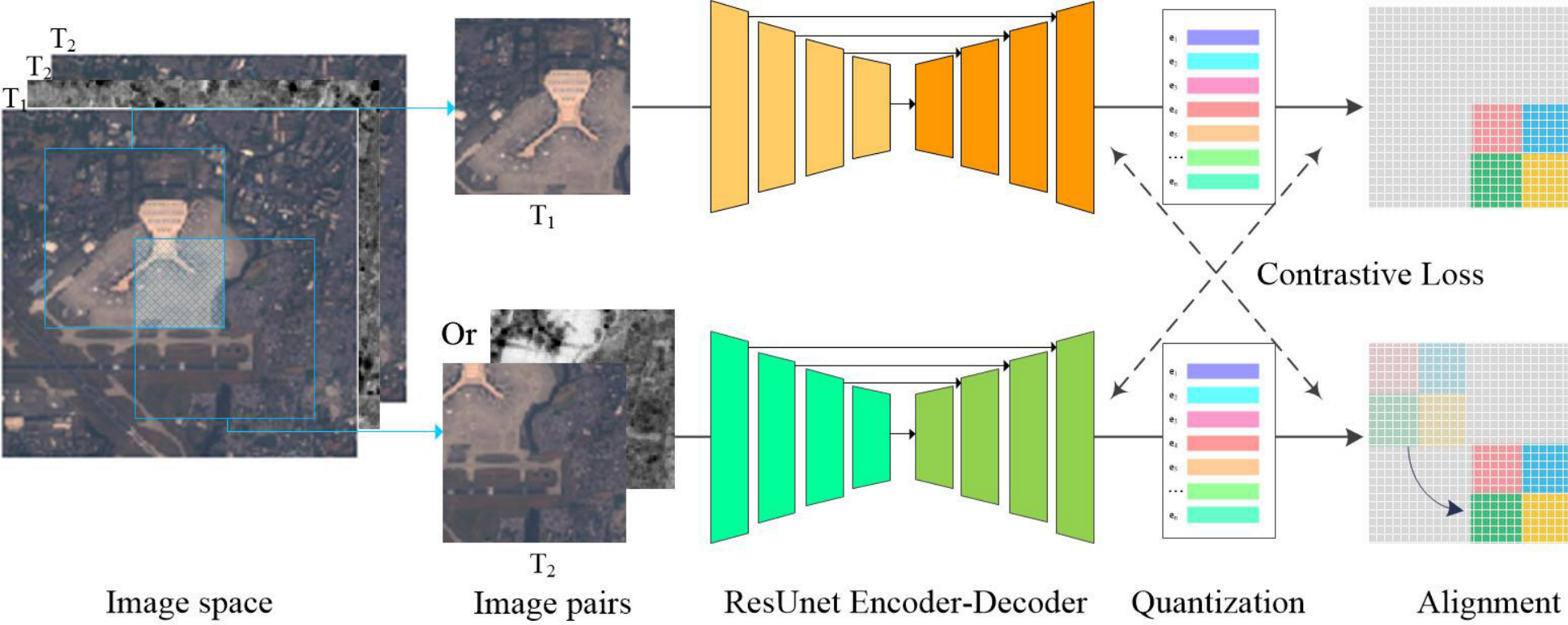}
	\caption{Overview of the proposed pixel-wise self-supervised change detection approach. An illustration of pixel-wise representation learning framework from shift equivariance. There is an offset between two input views but still keeps an overlap. The framework is a Siamese architecture where two branches both consist of a ResUnet block and a vector quantization. At the end of the network, the output features of the ResUnet block and quantization between two branches are used as the input in the contrastive loss. Noted that the vector quantization just on one branch with the optical images into at the heterogeneous change detection scenario.}
	\label{fig1}
\end{figure*}

\section{Methodology}
\subsection{Network Architecture}
The proposed approach has two branches (Fig. \ref{fig1}), where two views of images are used as inputs in each branch.
Each branch contains a ResUnet block ($U$) \cite{zhang2018road} followed by a vector quantization ($V$).
In addition, the same shift transformation performed on input and output follows the shift equivariance principle.
This is used as a kind of geometric data augmentation.
We adopt a similar ResUnet architecture as the \cite{zhang2018road}, but replacing all padding types with the same padding.
Like U-net, ResUnet consists of encoder, decoder, and skip connections while using residual units instead of plain neural units.
The encoder is used to get compact features with deconvolution layers, while the decoder reconstructs the compact features at the pixel level.
Multi-level features from the encoder are aggregated by using skip connections, which makes the network with fewer parameters but better performance.

As the deeper and deeper of networks, the gradient in backpropagation sometimes vanishes, which results in a degradation problem.
In this context, He \textit{et al.} \cite{he2016deep} proposed the deep residual network using skip connections in each residual unit.
Each residual unit can be illustrated as a general form:
\begin{equation}
\begin{array}{l}
\mathbf{x}_{i+1}=\mathbf{x}_{i}+\mathcal{F}\left(\mathbf{x}_{i}, \mathcal{W}_{i}\right) \\
\mathbf{x}_{i+1}=f\left(\mathbf{x}_{i+1}\right)
\end{array}
\end{equation}
where $\mathbf{x}_{i}$ and $\mathbf{x}_{i+1}$ are the input and output of the $i-th$ residual unit, $\mathcal{F}(\cdot)$ is the residual function, $f(\cdot)$ is activation function.

In this work, each residual unit of encoder consists of two BN-ReLU-Conv blocks and an identity mapping (Conv-BN). 
And, there are three residual units in the encoder and a Conv-BN-ReLU-Pool block before the residual units.
In the last two blocks, instead of using a pooling operation to downsample the feature map size, a stride of 2 is applied to the convolution block to reduce the feature map by half.
The decoder part has three blocks but without residual connections, which used a stride of 1 in all convolution.
In each decoder block, there is a concatenation with the feature maps from the corresponding encoding path and then an up-sampling operation for concatenated feature maps.
After the last level of the decoding path, a linear layer is used to reconstruct the learned representations.

We introduce a vector quantization module $V$ to augment learned representations of each branch.
The quantization module takes the pixel-wise representation $z_t$ from the ResUnet module, and map it to a new representation $v_t$.
This is done by selecting one entry from fixed-size codebook $C=\left\{c_{1}, \cdots, c_{N}\right\}$, where $N$ is the size of the codebook, and apply a linear transformation to obtain $v_t$.
However, this selecting process is not differentiable.
To mitigate the problem, we use the Gumbel-Softmax loss with reparameterization trick \cite{baevski2020wav2vec}.
In order to train which entry to select, we map the encoder output $z$ to logits  $\mathbf{l} \in \mathcal{R}^{N}$ via a linear layer, and the probability of selecting the $j-th$ code in the codebook is defined as follows:
\begin{equation}
p_{j}=\frac{\exp \left(\mathbf{l}_{j}+n_{j}\right) / \tau}{\sum_{k=1}^{N} \exp \left(\mathbf{l}_{k}+n_{k}\right) / \tau}
\end{equation}
where $\tau$ is a non-negative softmax temperature, $n=-log(-log(u))$ and $u$ are uniform samples from $\mu (0,1)$.
The index with the largest value in logits $l$ will be chosen during the inference.
During the forward pass, codeword $i$ is chosen by $i = argmax p_j$ and in the backward pass, the true gradient of the Gumbel softmax output is used.
In this way, the output feature of the projector is discretized to a finite set of representations with a vector quantization module.

Shift equivariance is achieved by using shift operation on the input image and output features of two branches individually.
Specifically, given an image pair, we randomly crop the same size area in each image and keep overlap between two cropped areas.
During the training, the cropped image pair is fed into two branches, respectively, to obtain two pixel-wise representations.
To align pixel-wise representations of two branches, the reverse transformation is applied to the outputs.
During the inference, the model provides two feature maps for the considered bi-temporal images.
The change intensity map is defined as the cosine similarity between the feature vectors of bi-temporal images.
To get binary change maps, the Rosin thresholding method is used in this work.

\subsection{Loss Function}
The training objective contains two parts.
One is a contrastive loss which is used to distinguish the representations of each pixel from others.
The loss is sampled over the pixels lied in the overlap areas between two branches.
This aims to keep the consistency of the normalized pixel-wise representations between two branches.
Each pixel-wise feature pair $(v_1^i,v_2^i)$ is sampled from the same location $i$ that is called positive.
Let $v_2^j$ be taken from another location that is called negative.
The contrastive loss can written as $\mathcal{L}_{\text {contrast}}$:
\begin{equation}\label{eq3}
\mathcal{L}_{\text {contrast}}=-\underset{S}{\mathbb{E}}{\left[\log \frac{h_{\theta}(v_1^i, v_2^i)}{\sum_{j=1}^{N} h_{\theta}(v_1^i, v_2^j)}\right]}
\end{equation}
where $h_\theta(\cdot)$ is a similarity function (i.e., cosine similarity), $(v_1^i,v_2^i)$ is the normalized latent representation of secen $i$, $(v_1^i,v_2^j|j\ge i)$ is the normalized latent representation of negative pair and $S=\{s_1^1, s_2^1,s_2^2, \cdots, s_2^{N-1}\}$ is a set that contains $N-1$ negative samples and one positive sample.
In the training process, the network is trained to increase the value of positive pairs and decrease the value of negative pairs.
This results in a feature representation that is close to positive pairs whereas it is not for negative pairs. 
Compared with the instance-level contrastive learning, this loss function is able to make the model get more detailed representations and is more suitable for dense prediction downstream tasks.

Together with the contrastive loss, a codebook loss \cite{baevski2020wav2vec} is designed to encourage the equal use of all entries in the codebook.
This is achieved by maximizing the entropy of the averaged softmax distribution $\bar{p}_{v}$ over the entries $V$ for each codebook in a batch.
Formally, the codebook loss is defined as:
\begin{equation}
\mathcal{L}_{d}= -\frac{1}{V} H(\bar{p}) = \frac{1}{V} \sum_{v=1}^{V} \bar{p}_{v} \log \bar{p}_{v}
\end{equation}
Finally, we use the pixel-wise contrastive loss in conjunction with the codebook loss, leading to the total loss:
\begin{equation}
\mathcal{L}_{\text {total}}=\mathcal{L}_{c} + \mathcal{L}_{d}
\end{equation}

\subsection{Uncertainty-aware Feature Learning}
In this work, we propose to use a deterministic model to approximates the feature representations invariant to the seasonal changes.
Specifically, the model learns to directly infer both feature representation and its uncertainty on one forward pass.
The network architecture is straightforward using a teacher-student paradigm similar to the knowledge distillation.
Given a pre-trained model as the teacher model that is trained by the proposed approach, bi-temporal predictive samples are generated from the teacher to train the student.
Like these works in regression task, we use the KL loss to approximate the variational predictive distribution and estimate the log variance ($s$) from network directly inveading the gradient explosion.
The loss function can be written as:
\begin{equation}
\mathcal{L}_{u}=\frac{1}{H} \frac{1}{W} \sum_{i} \frac{1}{2} \exp \left(-s_{i}\right)d({y}_{i}^m-{\mu}_{i}^n)+\frac{1}{2}{s}_{i}
\end{equation}
where $i$ corresponds to each pixel within an image; $H$ and $W$ are the height and width of the image; $y^m$ and ${\mu}^n$ are the predictions from teacher and student network at time $m$ and $n$, respectively.
In most works, $d$ is $L_2$ distance between the prediction from teacher and student network while we use cosine distance substituted for $L_2$ distance.
Empirically, we found that training solely with the above loss functions sometimes leads to sub-optimal predictive performance. 
This may be due to too much noise between different temporal images. 
Thus we leverage the features of the image at the same time $m$ from the teacher network to stabilize the training of the student model. 
The teacher-student model is trained with the consine distance of bi-temporal features, leading to the total loss:
\begin{equation}
\mathcal{L}_{\text{total}}=\mathcal{L}_{u} + \lambda \frac{1}{H} \frac{1}{W} \sum_{i} d\left({y}_{i}^m  -{\mu}_{i}^m\right)
\end{equation}
where the $\lambda$ is a hyper-parameter to be tuned. 
We found that $\lambda = 1$ generally performs well for our experiments.

\begin{figure}[pt]
	\centering
	\includegraphics[width=3.5 in]{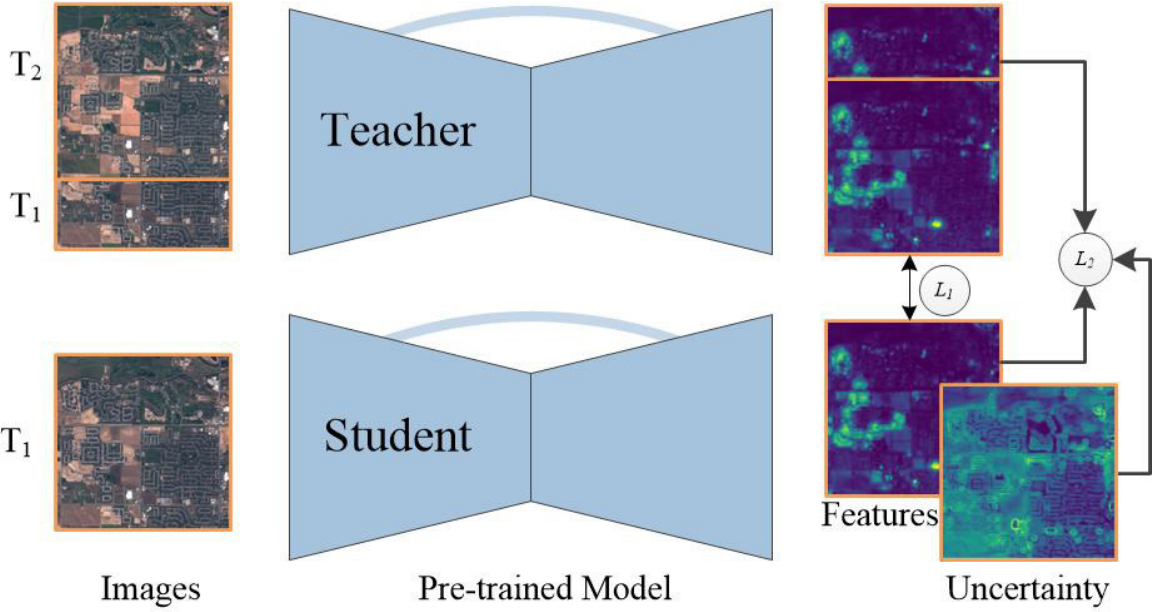}
	\caption{Overview of the Teacher-Student Paradigm for Uncertainty-aware Feature Learning.}
	\label{fig2}
\end{figure}

\section{Experiments}
In this section, we present the considered datasets at first and then evaluate the representations obtained using the proposed approach for binary change detection tasks.
\subsection{Description of Datasets}
We developed our experiments on three multi-view data sets, where two homogeneous data sets and one heterogeneous data set are used. 
\subsubsection{OSCD\_S2S2}
The Onera Satellite Change Detection (OSCD) dataset \cite{daudt2018urban} was created for bi-temporal change detection using Sentinel-2 images acquired between 2015 and 2018.
The dataset was acquired in 24 cities and including different landscapes.
The pixel-wise ground truth labels, which were manually annotated, focus on urban growth and built-up changes while containing some errors on the identification of bare lands.
To use this dataset in self-supervised training, we downloaded additional Sentinel-2 images in the same location as the original bi-temporal images between 2016 and 2020.
We considered images from each month to augment existing image pairs.
To keep consistency with previous research, 10 image pairs obtained from 10 different cities are treated as the test set for evaluation.

\subsubsection{MUDS\_S2S2}
Multi-temporal Urban Development (MUDS) dataset is an open-source dataset of the native Planet 4 m resolution imagery between 2017 and 2020.
The imagery comprises 24 consecutive monthly mosaic images of 101 locations over 6 continents.
To use this dataset with OSCD\_S2S2, we downloaded additional Sentinel-2 images in the same location as the original images between 2017 and 2020 and resized each image to 512 $\times$ 512 pixels.
We chose 30 of 110 locations as the evaluation set, where the first image was defined as pre-images and the last image was defined as post-image.
In addition, we manually labeled the three types of change such as built-up, bare land, and water.

\subsubsection{Flood in California}
The California dataset is a cross-sensor data set that includes a Landsat-8 (multi-spectral) and a Sentinel-1 GRD (SAR) image.
The multispectral and SAR images are acquired on 5 January 2017 and 18 February 2017, respectively.
The dataset represents a flood that occurred in Sacramento County, Yuba County, and Sutter County, California.
The ground truth was extracted from a Sentinel-1 SAR image pair where the pre-event image is acquired approximately at the same time as the Landsat-8 image \cite{chen2021self}.
The other three image pairs of Sentinel-1 and Landsat-8 images of the same scene acquired in 2017 and 2018, respectively, were used for the self-supervised pre-training of the proposed approach.

\subsection{Experiment Settings}
\subsubsection*{Evaluation Metrics}
To appraise the different methods in binary change detection, five evaluation metrics (precision (Pre), recall (Rec), overall accuracy (OA), F1 score and Cohen's kappa score (Kap)) are used in this paper.
We simply classify the image pixels into two classes by setting an appropriate threshold value according to the Rosin thresholding method and analyze them with reference to the ground truth map.

\subsubsection*{Implementation Details}
Concerning the data augmentation for different views, we further apply random flip to improve the performance of the proposed approach in the training of both methods.
The photometric augmentation was not considered, because we want to capture the seasonal change better.
We used 1024 codewords in each codebook which are utilized to approximate the quantized output, where the closest codeword contributes the most.
To select the appropriate samples for calculating contrastive loss, we segment the image into superpixels and select one point from each superpixel.
We used the felzenszwalb approach \cite{felzenszwalb2004efficient} to generate superpixels and gamma is et to 0.5, scale is set to 200.

For self-supervised training, we adopt Adam with an initial learning rate of $3e^{-4}$ and decay the learning rate with the step scheduling without restarts and set the batch size as 100.
Models are run for 1000 epochs.
We used bi-temporal images to train teacher-student paradigm for uncertainty-aware feature representation.
In order to capture the teacher predictive distribution, the image used to train the student model should not be the same as the one for the teacher.
To alleviate this problem, we used bi-temporal images during the training of the teacher-student network, both temporal images were input in the teacher model and one of them was input in the student model.
This extra image is crucial for the enhanced quality of uncertainty estimations.
We emphasize that the uncertainty estimates come from the bi-temporal images rather than this extra image.
The teacher model comes from self-supervised training.
To achieve faster convergence, we initialize the student network using the weights of the teacher network.
To this end, a smaller initial learning rate of $5\times 10 ^{-4}$ is used to train the student network for 1000 epochs.
We employ a step learning rate policy on the student network only and a batch size of 10.

There are two baseline approaches, patch-based self-supervised approach (PatchSSL) and auto-encoder based approach (CAA), that we categorize to make a comparison.
Besides these two approaches, we also used another SCCN as a comparison in heterogeneous change detection.
Meanwhile, we also include the results of the teacher model in the proposed approach as a baseline comparison.
Note that we only used vector quantization in the branch with the input of optical images in a heterogeneous setting.

\begin{table}[pt]
	\centering
	\caption{Quantitative evaluations of different approaches applied to the OSCD\_S2S2 dataset.}
	\label{table1}
	\renewcommand\tabcolsep{5.0pt}
	\centering
	\begin{tabular}{ccccccc}
		\hline
		Type & Method & Pre(\%) & Rec(\%) & OA(\%) & F1 & Kap \\ \hline
		\multirow{5}{*}{\rotatebox{90}{Unsup.}}
		& PixSSLs. & 58.80 & 41.54 & 95.38 & 0.49 & 0.46 \\
		& PixSSLt & 36.00 & 49.23 & 92.70 & 0.42 & 0.38 \\
		& PatchSSL & 40.44 & 69.10 & 93.00 & 0.51 & 0.48 \\
		& CAA  & 23.49 & 52.96 & 91.66 & 0.33 & 0.29 \\	\hline
		\multirow{2}{*}{\rotatebox{90}{Sup.}} & FC-EF & \textbf{55.34} & 39.48 & 95.13 & 0.46 & \textbf{0.44} \\ 
		& FC-EF-res & 54.97 & 38.39 & 95.10 & 0.45 & 0.43 \\ \hline
	\end{tabular}
\end{table}
\begin{figure*}[pb]
	\centering
	\includegraphics[width=7.0 in]{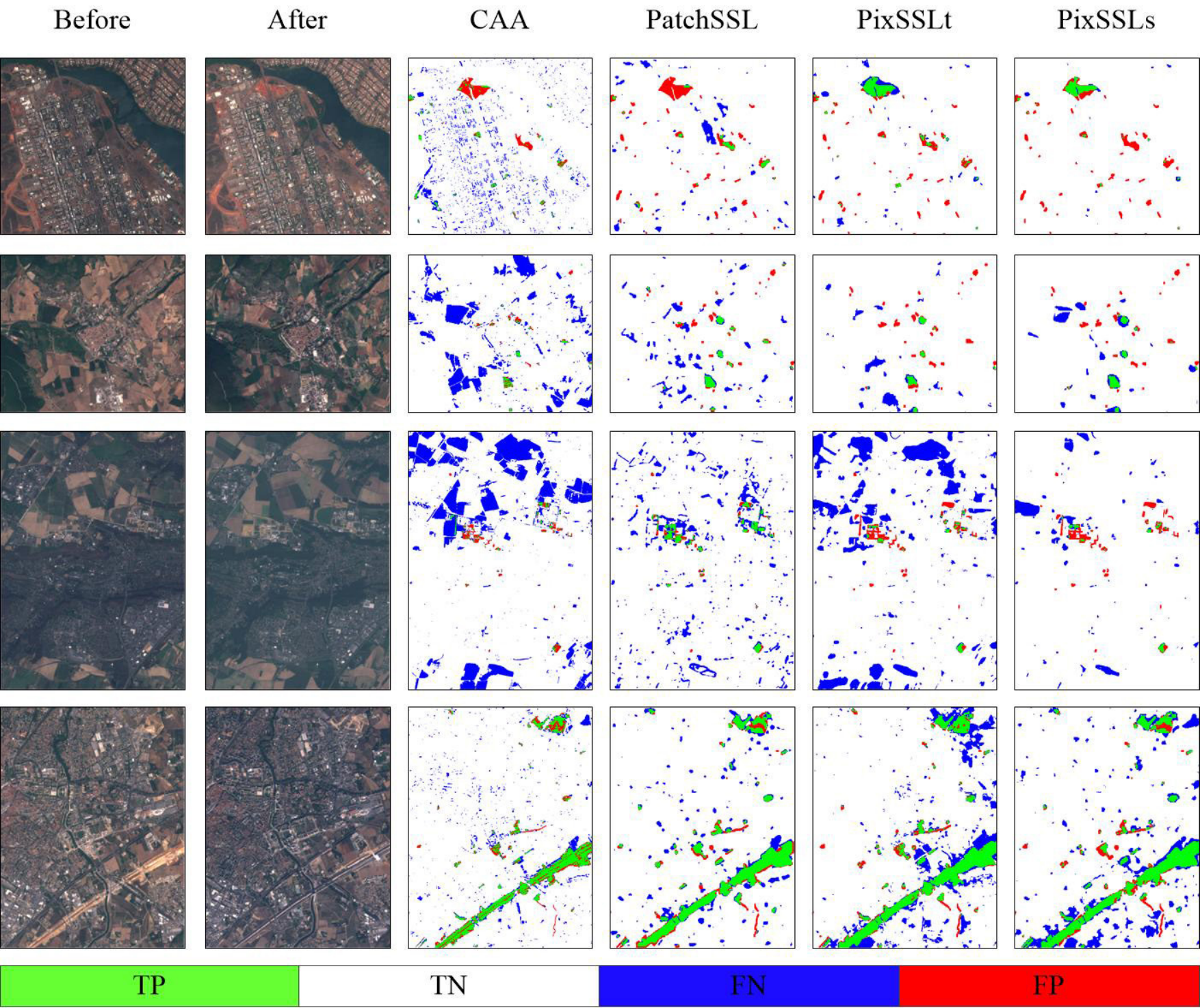}
	\caption{Examples of change detection results on OSCD\_S2S2, organized in one row for each location. Col. 1: pre-event image; Col. 2: post-event image. Change maps obtained by: CAA (Col. 3), PatchSSL (Col. 4), PixSSLt (Col. 4) and the PixSSLs (Col. 6). }
	\label{fig3}
\end{figure*}
\begin{figure*}[pb]
	\centering
	\includegraphics[width=7.0 in]{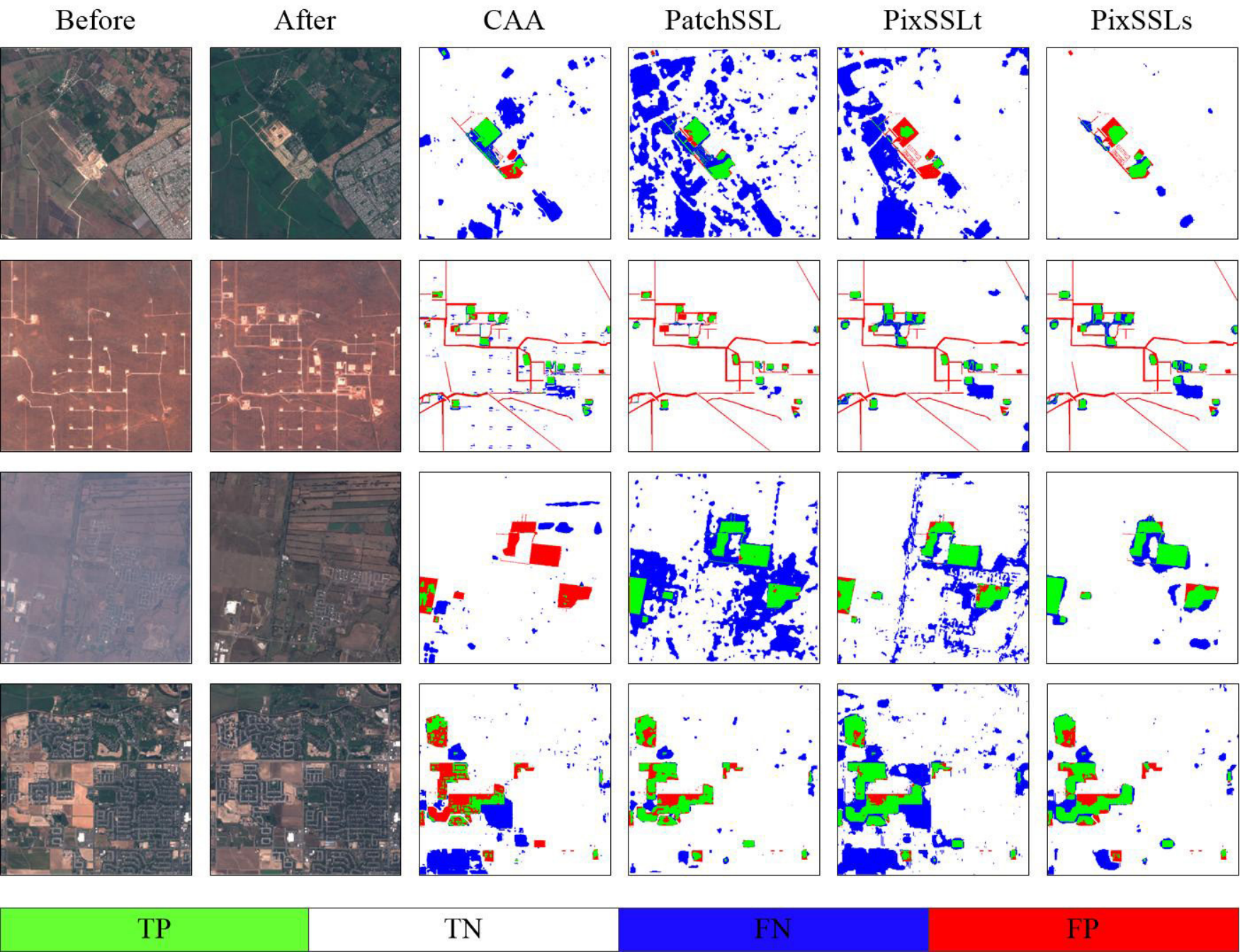}
	\caption{Examples of change detection results on MUDS\_S2S2, organized in one row for each location. Col. 1: pre-event image; Col. 2: post-event image. Change maps obtained by: CAA (Col. 3), PatchSSL (Col. 4), PixSSLt (Col. 4) and the PixSSLs (Col. 6).}
	\label{fig4}
\end{figure*}


\subsection{Experimental Results on bi-temporal Sentinel-2 images}
As mentioned in Section III, two multi-temporal Sentinel-2 images data set are proposed to evaluate the effectiveness of the proposed approach.
They include the OSCD\_S2S2 test set and MUDS\_S2S2 test set.
Performance of the proposed approach (PixSSLs) is compared with Code-Aligned Autoencoders (CAA), patch-based self-supervised methods (PatchSSL) and the result of the teacher model (PixSSLt) in the proposed approach.
The supervised results from previous research are also included in the OSCD data set.

The performance metrics obtained on the OSCD test set are reported in Table \ref{table1}.
As one can see that PixSSLt acquired an OA of 92.70 \% and a Kappa coefficient of 0.38, which outperforms the results obtained from CAA.
It also can be observed there is a big improvement on OA and Kappa with about 3\% and 0.12 after applying the uncertainty approach.
The results obtained from the PixSSLs used uncertainty approach are significantly better than those obtained using pixel-wise self-supervised approach only in almost all cases except the case of Recall.
Nevertheless, the patch-based self-supervised approach with an OA of 93.00 \% and a Kappa coefficient of 0.48, obtained the best performance on both F1 score and Kappa.
We also introduced the results of the supervised approach as presented in \cite{chen2021self}.
Noted that the proposed PixSSLs and the PatchSSL approaches even outperform the supervised approach in this dataset.
In general, PatchSSL gets a similar result as the uncertainty-enhanced PixSSLs.
While they reach comparable performance, the PixSSLs is more efficient in application.

\begin{table}[pt]
	\centering
	\caption{Quantitative evaluations of different approaches applied to the MUDS dataset.}
	\label{table2}
	\renewcommand\tabcolsep{7.0pt}
	\centering
	\begin{tabular}{cccccc}
		\hline
		Methods & Pre(\%) & Rec(\%) & OA(\%) & F1 & Kappa \\ \hline
		PixSSLs & 39.84 & 67.93 & 94.26 & 0.50 & 0.47 \\
		PixSSLt & 29.58 & 65.00 & 92.04 & 0.40 & 0.37 \\
		PatchSSL & 31.40 & 67.59 & 91.18 & 0.43 & 0.39 \\
		CAA & 33.34 & 49.12 & 92.95 & 0.40 & 0.36 \\ \hline
	\end{tabular}
\end{table}

Similar performance can be found on the MUDS test set (Table \ref{table2}).
The PatchSSLt acquired a similar result as the PatchSSL, which outperformed the results of CAA.
Compared with the results on OSCD, the improvement is more prominent when using the uncertainty approach on pixel-wise contrastive approach.
Among all five performance indices, the results obtained using PixSSLs are better than those obtained using PixSSLt only in most indices except the case of precision with about 10\% drops.
Not only that, it even outperformed the PatchSSL which obtained an OA of 91.18\% and a Kappa coefficient of 0.39.
The PatchSSL performs relatively poorly, mostly due to the fact that the MUDS dataset with more seasonal changes.
On the contrary, the more built-up changes are contained in the OSCD dataset.

Besides the quantitative analysis, we also provide a visual qualitative comparison, where the TP, TN, FN and FP pixels are colored in green, white, blue and red, respectively.
In Fig. \ref{fig3}, we show a comparison between all methods on the OSCD test set.
One can see that the change maps obtained by CAA seem to be noisy and with more false alarms.
Instead, the change maps obtained by other methods are in general more accurate and less noisy.
Change maps provided by PixSSLt with more false alarms where plenty of unchanged pixels are wrongly classified as changed ones.
The uncertainty enhanced PixSSLs suppresses most unchanged regions but also fails to highlight some clearly changed regions.
Among all comparison methods, PatchSSL successfully detects most of the changed pixels and achieved the best change map. 
Although the PatchSSL still achieved the best change map, the proposed approach achieves a comparable result and is more efficient.

Fig. \ref{fig4} shows a comparison between all methods on the MUDS test set.
One can see that the results obtained by PatchSSL and PixSSLt failed to suppress most seasonal changes.
The results obtained by CAA perform better, while it contains lots of missing detection.
This issue is well addressed in the uncertainty-enhanced PixSSLs.
As shown in the first and third row, the PixSSLs successfully suppress most seasonal changed areas while the compared methods fail.
In contrast, most of the changed pixels are correctly detected in all contrastive approaches.
In general, the experiments on two datasets demonstrate that self-supervised methods obtained the best quantitative and qualitative performance with respect to the considered autoencoder approach and the uncertainty-enhanced PixSSLs give a big improvement in suppressing seasonal changes.

\begin{table}[pb]
	\centering
	\caption{Quantitative evaluations of different approaches applied to the Flood dataset.}
	\label{table3}
	\renewcommand\tabcolsep{7.0pt}
	\centering
	\begin{tabular}{cccccc}
		\hline
		Methods & Pre(\%) & Rec(\%) & OA(\%) & F1 & Kappa \\ \hline
		PixSSL & 41.64 & 71.14 & 90.51 & 0.53 & 0.48 \\
		PatchSSL & 40.43 & \textbf{68.14} & 90.24 & 0.51 & 0.46 \\
		CAA & \textbf{76.49} & 40.38 & \textbf{94.68} & 0.53 & 0.50 \\
		SCCN & 51.42 & 64.44 & 92.88 & \textbf{0.57} & \textbf{0.53} \\ \hline
	\end{tabular}
\end{table}

\begin{figure*}[pb]
	\centering
	\includegraphics[width=6.5in]{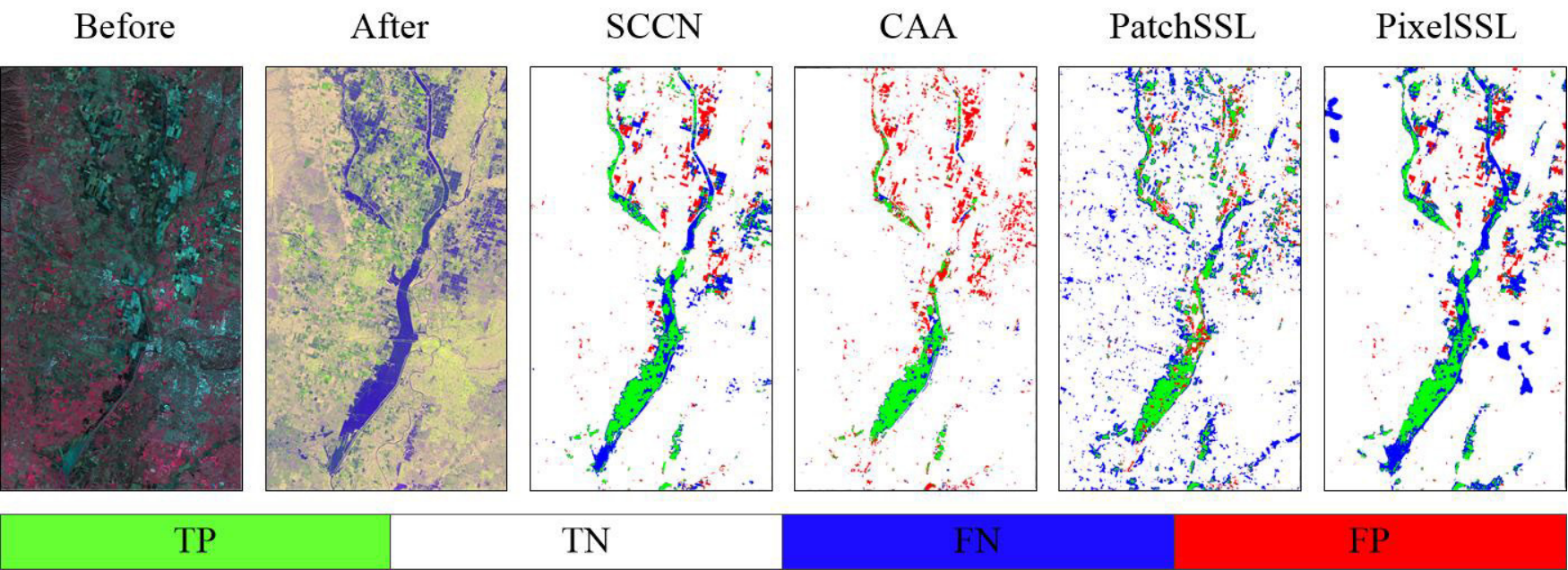}
	\caption{Change detection results on califorlia flood, organized in one row for each location. Col. 1: pre-event image (Landsat-8); Col. 2: post-event image (Sentine-1). Change maps obtained by: SCCN (Col. 3), CAA (Col. 4), and PatchSSL (Col. 5) and PixSSLt (Col. 6). }
	\label{fig5}
\end{figure*}

\subsection{Experimental Results on the multi-sensor image pair}
In the second change detection scenario, we consider one cross-sensor data set which consists of a Sentinel-1/Landsat-8 image pair (California flood).
Table \ref{table3} lists the quantitative statistics on the changed maps by four unsupervised methods (SCCN, CAA, PatchSSL and PixSSLt).
In this table, SCCN achieves the best F1 score, Kappa and the second-best values on Precision, Recall and OA, while the PixSSLt achieved the highest Recall.
The proposed PixSSLt and CAA have similar performance on F1 score and get the second-best performance.
Although PatchSSL and PixSSLt do not get the best performance, it is also very close to the other two results.

Fig. \ref{fig5} illustrates the Landsat 8 and Sentinel-1 images and the change maps from the compared methods.
SCCN achieves the best change maps with a clear boundary of flood areas, while the results obtained by CAA just detected the main flood area and missed the small areas.
PtachSSL highlights most of the flood areas, while it seems noisy and with more false alarms.
Compared with the PatchSSL, the proposed PixSSLt produced a more clear change map, while it still contains lots of false alarms.
The quantitative and qualitative analysis demonstrates the effectiveness of the proposed PixSSLt on multi-sensor change detection scenarios.

\subsection{Discussion}
\begin{figure*}[pt]
	\centering
	\includegraphics[width=7.0 in]{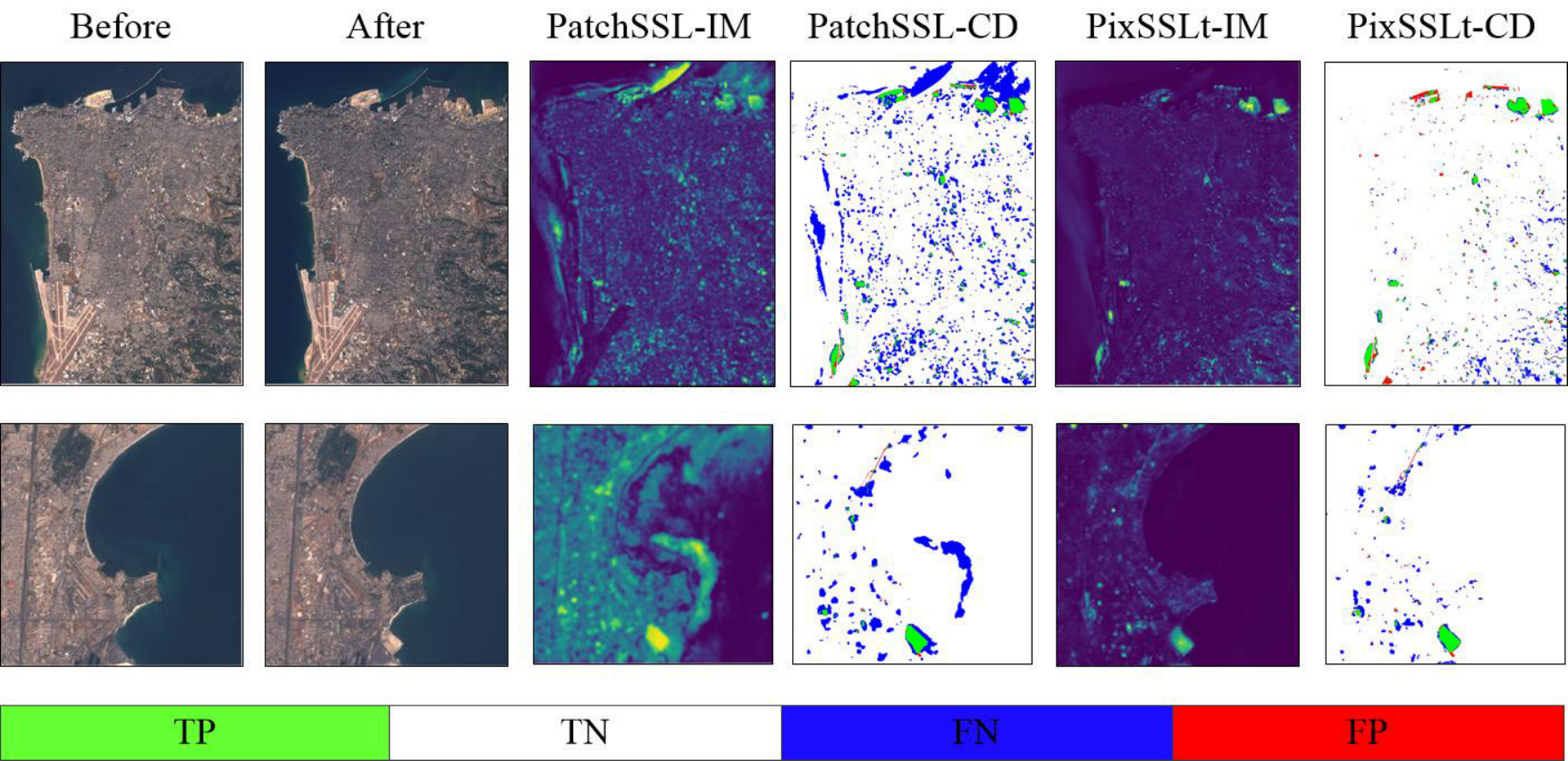}
	\caption{Examples of change intensity maps and change detection maps obtained by PatchSSL and PixSSLt on water area, organized in one row for each location. Col. 1: pre-event image; Col. 2: post-event image. Change maps obtained by: PatchSSL (Col. 4) and PixSSLt (Col. 6), and change intensity maps obtained by PatchSSL (Col. 3) and PixSSLt (Col. 5).}
	\label{fig6}
\end{figure*}

To better analyze the robustness of the proposed approach, we further evaluated the performance in terms of the five metrics considered under more challenging water areas.
Here we provide two examples: one comes from OSCD and the other from MUDS.
Fig. \ref{fig6} shows that the change intensity maps and change detection maps obtained by PatchSSL and PixSSLt.
As we can see, both scenarios present many false alarms of water areas using PatchSSL whares the results of PixSSLt correctly identified the water areas as non-change.
In a quantitative way, the proposed PixSSLt with an OA of 92.64 \% and a Kappa coefficient of 0.39, obtained the best performance on all five metrics.
This demonstrates again that the proposed PixSSLt is more robust in the complex scenario.

\begin{table}[pb]
	\centering
	\caption{Efficiency comparisons between different methods.}
	\label{table4}
	\renewcommand\tabcolsep{11.0pt}
	\begin{tabular}{cccc}
		\hline
		Models & Kappa & MACs (G) & Params (M) \\ \hline
		PixSSL & 0.50 & 84.9 & 4.216 \\
		PatchSSL & 0.48 & 8026.1 & 21.353 \\
		CAA & 0.29 & 40.7 & 0.103 \\ \hline
		FC-EF & 0.44 & 14.4 & 1.351 \\
		FC-EF-res & 0.43 & 8.1 & 1.104 \\ \hline
	\end{tabular}
\end{table}

\begin{table}[pt]
	\centering
	\caption{Change detection results of PixSSLs on the MUDS dataset using Rosin and otsu threholding methods.}
	\label{table5}
	\renewcommand\tabcolsep{5.0pt}
	\begin{tabular}{ccccccc}
		\hline
		Thresholding & Method & Pre(\%) & Rec(\%) & OA(\%) & F1 & kappa \\ \hline
		\multirow{2}{*}{Rosin} & PatchSSL & 31.40 & 67.59 & 91.18 & 0.43 & 0.39 \\
		& PixSSLs & 39.84 & 67.93 & 94.26 & 0.50 & 0.47 \\ \hline
		\multirow{2}{*}{OTSU} & PatchSSL & 18.43 & 87.06 & 80.49 & 0.30 & 0.24 \\
		& PixSSLs & 24.22 & 75.26 & 88.91 & 0.37 & 0.32 \\ \hline
	\end{tabular}
\end{table}

In order to have an intuitive understanding of the efficiency between different methods. 
We present a detailed comparison as follows (Table \ref{table4}). 
MACs and the number of parameters are considered as two relevant metrics of the model’s efficiency. 
The number of multiply-accumulate (MAC) operations is used to measure the computational cost. 
Following the common practice, we use them to measure the network efficiency, in terms of computational cost and memory consumption.
The metrics reported for all models are based on the use of PyTorch on a 7.8 GB RTX 2070ti GPU.

Table \ref{table4} shows the Kappa, MACs, the number of parameters and the inference time of each model. 
The performance of each model on Kappa is obtained on the OSCD test set.
From the analysis of Table \ref{table4}, one can see that CAA need much lower MACs but result in far low accuracy.
Unlike CAA, two self-supervised methods are heavy-weight networks and with high accuracy.
Compare with PatchSSL, the proposed PixSSL achieved competitive results but with much lower computational costs.

We then derived change maps with different thresholding methods (OTSU and Rosin) using two self-supervised methods: PixSSLs and PatchSSL. 
In Table \ref{table5}, we present the change detection results of the MUDS dataset utilizing the Rosin and OTSU thresholding method considering two self-supervised approaches.
The results of both methods obtained by Rosin are much better than their results using OTSU.
This indicates that the Rosin thresholding method is robust to the OSTU on the presented self-supervised change detection scenario.

\section{Conclusion}
In this work, we have presented a pixel-wise contrastive learning approach for multi-view remote sensing image change detection, which uses ResUnet and vector quantization as the architecture of the network.
The main idea of the presented approach is the use of contrastive loss in pixel-level features and the suppression of seasonal changes using the uncertainty method.

Experimental results on the multi-view remote sensing image dataset demonstrated its superiority and efficiency over the other state-of-the-art methods.
Results also show that uncertainty enhanced approach leads to a significant performance improvement with respect to the use of contrastive learning only.


\ifCLASSOPTIONcaptionsoff
  \newpage
\fi

\bibliographystyle{IEEEtran}
\bibliography{mylib}

\end{document}